# Effect of intrinsic local distortions vs. dynamic motions on the stability and band gaps ofcubic oxide and halide perovskites

Xin-Gang Zhao,[1] Zhi Wang,[1,2] Oleksandr I. Malyi,[1] Alex Zunger,[1,*]

[1] Renewable and Sustainable Energy Institute, University of Colorado, Boulder, Colorado 80309, United States
[2] State Key Laboratory for Superlattices and Microstructures, Institute of Semiconductors, Chinese Academy of Sciences, Beijing, 100083, China

* Corresponding author: Alex Zunger, (alex.zunger@colorado.edu; alex.zunger@gmail.com)

Ternary $ABX_3$ perovskites made of corner-sharing $BX_6$ octahedra have long featured prominently in solid-state chemistry and condensed matter physics. Still, the joint understanding of their two main subgroups—halides and oxides—has not been fully developed. Indeed, unlike the case in simpler compounds having a single, robust repeated motif ("monomorphous"), certain cubic perovskites can manifest a non-thermal (=intrinsic) distribution of local motifs ("polymorphous networks"). Such intrinsic deformations can include positional degrees of freedom (e.g., atomic displacements and octahedral tilting) or magnetic moment degrees of freedom in paramagnets. Unlike thermal motion, such intrinsic distortions do not time-average to zero, being an expression of the intrinsic symmetry breaking preference of the chemical bonding. The present study compares electronic structure features of oxide and halide perovskites starting from the intrinsic polymorphous distribution of motifs described by Density Functional Theory (DFT) minimization of the internal energy, continuing to finite temperature thermal disorder modeled via finite temperature DFT molecular dynamics. We find that (i) different oxide vs. halide $ABX_3$ compounds adopt different energy-lowering symmetry-breaking modes. The calculated pair distribution function (PDF) of $SrTiO_3$ from the first-principles agrees with the recently measured PDF. (ii) In both oxides and halides, such intrinsic distortions lead to band gap blueshifts with respect to undistorted cubic Pm-3m structure. (iii) For oxide perovskites, high-temperature molecular dynamics simulations initiated from the intrinsically distorted polymorphous structures reveal that the thermally-induced distortions can lead to a band gap redshift. (iv) In contrast, for cubic halide perovskite $CsPbI_3$, both the intrinsic distortions and the thermal distortions contribute in tandem to band gap blueshift, the former, intrinsic effect being dominant. (v) In the oxide $SrTiO_3$ and $CaTiO_3$ (but not in halide) perovskites, octahedral tilting leads to the emergence of a distinct Γ–Γ direct band gap component as a secondary valley minimum to the well-known indirect R–Γ gap. Understanding such intrinsic vs. thermal effects on oxide vs. halide perovskites holds the potential for designing target electronic properties.

## I. Introduction

ABX$_3$ halide A$^{(I)}$B$^{(IV)}$X$_3^{(VII)}$ and oxide ABO$_3$ perovskite often crystallize as corner-sharing low-symmetry (e.g., orthorhombic or tetragonal) phases at low temperatures (LT), sometimes transforming at higher temperatures into the cubic phase. They are generally discussed in terms of the nature of the microscopic degree of freedom, be that the magnetic moments (in ferromagnet/paramagnet), or electric dipoles (in ferroelectric/paraelectric), or octahedral degrees of freedom (e.g., tilting and rotations) [1–5]. X-ray diffraction (XRD) measurements [6] on such cubic perovskites have traditionally been modeled by the nominal cubic (Pm-3m) space group structure containing a single ABX$_3$ formula unit, with undistorted octahedra, where atoms reside on the nominal Wyckoff positions [7].

Such single-motif (monomorphous) cubic (Pm-3m) models of numerous oxide and halide perovskites [8–10] have been extensively used over the years in the theoretical literatures as input to the electronic band structure [11-19] and phonon lattice dynamics [20-26], constituting the basis for the currently accepted interpretation of a broad range of electronic properties [27-38].

However, theoretical predictions of electronic and structural properties based on such nominal cubic models do not always agree with experimental observations on the corresponding pristine bulk compounds. Puzzling contradictions include the observation of effects commonly known to be enabled by non-cubic symmetry, yet seen in nominally cubic phases, such as piezoelectricity in paraelectric BaTiO$_3$[33], Rashba effect in CsPbCl$_3$ [34], and second harmonic generation (SHG) in BaTiO$_3$ [35]. Furthermore, cubic models for phonon dispersion calculations often predicted dynamically unstable phonons in BaOsO$_3$, SrMnO$_3$, and SrFeO$_3$ [36,37]. In addition to these qualitative failures, quantitative disagreements were noticed between electronic structure calculations based on the nominal cubic structure and measurements, such as the predicted absence of effective mass enhancement in SrVO$_3$, BaTiO$_3$, LaMnO$_3$, and SrBiO$_3$ [29], underestimated band gaps not only for halide perovskite insulators such as CsSnI$_3$, CsPbI$_3$ and CsPbBr$_3$ [28,38], but more dramatically, the prediction of metallic electronic structures for known (Mott) insulators such as CaMnO$_3$, LaFeO$_3$, LaVO$_3$, YNiO$_3$, LaTiO$_3$, and YTiO$_3$ [30-32].

The views of two communities on the nature of the para- phases: Two leading communities—the "phase transition community" and the "electronic structure community"—are grappled with the understanding of the phase transition between the long-range ordered (LRO) low-temperature phases (e.g., orthorhombic, monoclinic, and tetragonal) and the high-temperature para-phases those lack LRO.

The "phase transition community" focused initially on two pictures: (1A) The displacive/soft-mode picture emphasizing time averaging, thermal displacements, and reciprocal space treatments of phonons, based on the monomorphous, high-symmetric minimal cell such as the cubic (Pm-3m) structure. A key feature of assuming a monomorphous structure is that all intrinsic distortions are thought to be negligible, and atoms are oscillating thermally around their Wyckoff positions. Therefore, one can always apply the time-average to the observable properties. Reciprocal-space calculation of phonons is applied, often leading to the softening of the specific optic phonon (such as the F$_{1u}$ mode) as the transition to the lower-temperature low-symmetry phase is approached, where this mode is replaced by the lower symmetry modes of the LT phase. The displacive model was supported by early theoretical studies [39-41] that predicted no atomic displacements in the high-symmetry phases and no phase transition until strains were induced.

However, it was later realized experimentally that some crystals show intrinsic symmetry breaking in the high-symmetry para-phases. This approach is known as the (1B) order-disorder picture, which emphasizes spatial averaging, intrinsic displacements, and a real-space treatment of the ordering process. Examples include the multi-site off-center displacement model [33,40-43] in ferroelectric perovskites. The phonon modes obtained in such systems could include modes that do not soften, such as the "central mode" representing jumps between potential wells. In contrast, in the displacive view, the system oscillates in a single well creating a soft mode [44].

The "electronic structure community" grappled with the question of how to represent the microscopic atomic-scale structure of the para-phase needed for both phonon and band structure calculations. The properties of interest were not restricted to phonon, but included optical properties, transport (e.g., metal vs. insulator), etc. Early approximations involved using a monomorphous model, representing the average atomic structure (e.g., a non-electric representation of a paraelectric phase) or average spin structure (e.g., a non-magnetic representation of a paramagnet). Such monomorphous descriptions of the para-phases end up in electronic structure calculations with an artificially high symmetry (virtual-crystal-like), e.g., centrosymmetric (here, Pm-3m) structure. Such calculations for cubic oxide [10,45] and cubic halide [38] perovskites have been shown to give significant contradictions with the experiment. For $d$-electron paramagnets such as $LaTiO_3$ [46,47], a monomorphous model resulted in the metallic band structure, even though they are known to be (Mott) insulators. In cubic halides, the predicted band gap values were anomalously underestimated [38]. Such broad disagreements between experiments and calculations based on the monomorphous approach were generally addressed in the "electronic structure community" by two diverging approaches: In approach 2A (dynamic correlation in symmetry unbroken structures), the use of simple periodic monomorphous cell persisted. In the cases of oxide perovskites, the predicted false metallic states in $d$-electron cubic perovskites [31,32] and the absence of mass enhancement [29] motivated, in part, the development of the "theory of strongly correlated compounds" [48]. In such approach, the nominal high-symmetry monomorphous structures were retained. But strong interelectronic correlations codified by the Mott-Hubbard electron repulsion U were invoked in the symmetry unbroken (monomorphous) structures to reconcile such predicted electronic discrepancies with measurements. This was addressed by introducing time fluctuation in the microscopic dipole, magnetic moments, geometric tilting, which could lead to an average zero polarization (in the paraelectric phase) or an average zero magnetic moment (in the paramagnetic phase). The second approach (2B) is the mean field theory in symmetry broken structures, such as Density Functional Theory (DFT).

Initially, in this approach, the disorder in the para-phases was treated by "structural approximants" avoiding molecular dynamics (MD). Notable examples include the "Virtual Crystal Approximation (VCA) [49] and "Coherent Potential Approximation" (CPA) [50]. These approaches offer an effective configuration that could be used as input to standard electronic structure theory and capture some aspects of the disorder. Whereas such models are ubiquitous for positionally disordered alloys, they are not generally applicable for crystallographically ordered single-phase compounds. The discrepancies above between such simple monomorphous models and experiments are, therefore, a persistent difficulty. Approach (2B) then abandoned the simple VCA and CPA and further abandoned the monomorphous restriction by using instead supercell that affords spatial symmetry breaking. It was pointed out that intrinsic symmetry-breaking can occur spontaneously if no artificial (e.g., monomorphous) restrictions were placed. This intrinsic symmetry-breaking leads to the emergence of distribution of local motifs in the form of a distribution of non-zero magnetic moments (in paramagnets), or electric dipoles (in paraelectrics), or atomic displacements (in paraelastic), or a mixture of such microscopic degrees of freedom [27-30]. Furthermore, such a formation of a polymorphous network is accompanied by a lowering of the internal energy. [The term polymorphous network refers to different local motifs (i.e., presenting microscopically different distortions off Wyckoff position, hereafter DOWPs) in different local environments) in the same phase instead of different local motifs in different phases. The latter is known as polymorphs].

This paper will address the difficulties mentioned above in oxide and halide cubic perovskites by investigating the different local environments both intrinsically and via thermal effect, replacing at the outset the monomorphous approach by the polymorphous extension. The starting point of this discussion is that the tradition of seeking the smallest possible, highest symmetry unit cell is not protected by fundamental physical principles and may need to be relaxed under some circumstances. Indeed, the real cubic perovskite phase may manifest a distribution of positional degrees of freedom (e.g., displacements and tilting [51]), or magnetic moment degrees of freedom (as in paramagnets [52]), or local dipole degrees of freedom (as in paraelectrics [33]) even in single phase compounds. It was pointed out [27-30] that unlike the case in simpler compounds having a single, robust repeated motif intrinsically (e.g., the tetrahedron in covalent semiconductors), the chemical bonding in certain cubic perovskites creates a intrinsic distribution of energy lowering local motifs referred to as a polymorphous network [28,29]. The driving forces for stabilizing such symmetry breaking are varied, and can include lone pair electrons (in B2+ = Pb, Sn, etc.) or Jahn-Teller ions, leading to intrinsic atomic DOWPs. Such polymorphous networks manifest a lower calculated internal energy $U_0$ of the non-vibrating lattice relative to the monomorphous network. This energy lowering associated with the formation of intrinsic DOWPs can be evaluated by constraining the global shape of the structure to the cubic shape and minimizing the internal energy with respect to all cell-internal (atomic positions, e.g., octahedral tilting and ferroelectric displacements) degrees of freedom [25,26,28]. Such intrinsic DOWPs do not time-average to zero, as thermal motions do, being an expression of the intrinsic symmetry breaking preference of the chemical bonding.

## II. The type of systems discussed in the current symmetry-breaking approach: Intrinsic and dynamic

We will discuss two types of distortions: First, single-mode distortions are isolated DOWPs such as single Glazer mode, B site off-center, Jahn-Teller distortion, and breathing mode. These single-mode distortions are explored in 2x2x2 supercell by freezing in a particular distortion mode, one at the time of a given amplitude, keeping the structure frozen. This is done for the general understanding of how specific compounds respond to given symmetries of single-mode DOWPs. Second, our primary calculation of polymorphous structure involves allowing for multi-mode DOWPs in a cubically shaped 4x4x4 supercell, where all atoms are relaxed by nudging their positions and following the Hellman-Feynman forces.

Table 1 provides an overview of the general trends in DFT results of how intrinsic multi-mode and dynamic distortions affect band gaps in two prototype cubic perovskites, $SrTiO_3$ and $CaTiO_3$. The reason for choosing these systems is that we plan to study the effects on stabilities and band gap values due to the distortions related to different Goldschmidt factor and band-edge states. In this paper, we mainly focus on the semiconductor oxide $ABO_3$ (A = Ba, Sr, Ca, Pb; B = Ti, Zr), and halide $CsBX_3$ (B = Ge, Sn, Pb; X = F, Br, I), $KPbI_3$ and $RbPbI_3$, and show three structural depictions of the cubic phases: (a) the highly symmetric nominal cubic (Pm-3m) structure that prohibits distortions, (b) the intrinsic, non-thermal multi-mode distortions characteristic of the polymorphous network obtained through T = 0 energy minimization of DFT cubic supercell, and (c) the structure obtained at finite temperature from DFT molecular dynamics. The main features we wish to discuss are:

(1) Intrinsic multi-mode distortions led to energy lowering symmetry breaking generally associated with blueshifts in the band gaps: The (a) nominal monomorphous cubic structure excludes by symmetry possible DOWPs, leading in electronic band structure calculations to underestimated band gaps. At the extreme, in oxides with *d* electron B atoms, monomorphous models of paramagnets lead to false metallic states [31]. (b) Intrinsic distortions allowed in larger cubic supercells result in band gap blueshift in polymorphous cubic oxide perovskites $ATiO_3$ (A = Ba, Sr, Ca). We mention in passing that in the Mott oxides with *d* electron B atoms, the formation of a distribution of local moments in the paramagnetic cubic phase was recently shown [28] to correctly convert false metals in the monomorphous nominal cubic case to insulators with effective mass enhancement effect [29] in the polymorphous cubic phase without invoking dynamic correlation effects. These effects of intrinsic DOWPs can be measured using techniques sensitive to the local structure, such as atomic PDF and X-ray absorption fine structure (XAFS) [54,55]. The polymorphous nature in cubic $SrTiO_3$ is consistent with the agreement found here between our DFT calculated PDF of polymorphous $SrTiO_3$ and experimental measurement. For halide perovskites such as $CsPbI_3$, the polymorphous network corrects the quantitative underestimation of the calculated band gap relative to the experiment [38]. Band unfolding of the (supercell) band structure of $SrTiO_3$ and $CaTiO_3$ reveals that the energy lowering symmetry breaking results in the emergence of a direct band gap Γ–Γ component in addition to the indirect R-to-Γ transition.

(2) Dynamic distortions akin to stochastic thermal motion can lead to redshifts in the band gaps: The intrinsic DOWPs can be viewed as the low-temperature kernel of the thermal displacements associated with the non-vibrating free energy F = $U_0$–TS. The main differences between intrinsic DOWPs and dynamic thermal distortions are that, unlike the thermal displacements whose time averages represent the equilibrium structure with atoms on the nominal Wyckoff positions, the intrinsic DOWPs do not time average into the ideal Wyckoff positions. Large amplitude distortions in cubic oxide perovskites (octahedral tilting angle >8°) were found to turn over the band gap blueshift to a redshift. Indeed, the enhanced distortions seen in our MD simulations at high temperatures for oxides result in a significant band gap redshift (0.5–0.8 eV) in cubic $CaTiO_3$ and $SrTiO_3$ with respect to undistorted structure (Table 1). The thermal motion could eventually decrease the band gap at high temperatures towards the limit of metallization, as is the case in $BiFeO_3$ [56]. In contrast, for cubic halide perovskite $CsPbI_3$, the intrinsic distortions in the polymorphous network already give a large blueshift, and the thermal effects add another (small) blueshift. This finding corresponds to an anomalously low thermal gap renormalization, whereas using the monomorphous reference points predicts a spuriously large thermal blueshift [38].

**Table I.** Overview of density functional theory results discussed in the present paper of atomic distortions and ensuing calculated band gaps in three renderings of the cubic perovskite structure for $SrTiO_3$ and $CaTiO_3$: (a) nominal cubic Pm-3m model disallowing atomic distortions reduces band gap energies; (b) supercell model allowing intrinsic, energy lowering multi-mode distortions raises the band gap, and (c) dynamic thermal motion captured by molecular dynamics reduces the band gap at high temperature (5x5x5 $SrTiO_3$ supercell at 1400 K [53]; 4x4x4 $CaTiO_3$ supercell at 2000 K).

| Compounds | Property | Configurations of cubic $SrTiO_3$ and $CaTiO_3$ | | |
|---|---|---|---|---|
| | | (a) No distortion | (b) Intrinsic distortion | (c) Thermal distortion |
| $SrTiO_3$ | Sr displacements (Å) | 0.00 | 0.02-0.04 | 0.00-0.72 |
| | Ti displacements (Å) | 0.00 | 0.06-0.08 | 0.00-0.57 |
| | $TiO_6$ tilting angle (°) | 0.00 | 0-7 | 0-19 |
| | Band gap (eV) | 1.63 | 1.86 | 1.14 |
| $CaTiO_3$ | Ca displacements (Å) | 0.00 | 0.00-0.08 | 0.01-0.80 |
| | Ti displacements (Å) | 0.00 | 0.01-0.12 | 0.01-0.77 |
| | $TiO_6$ tilting angle (°) | 0.00 | 1-15 | 0-26 |
| | Band gap (eV) | 1.70 | 2.08 | 0.93 |

### III. The choice of supercell size matters

The physically meaningful cell used to express diffraction data might be larger than a macroscopically averaged minimal cell. As Fig. 1 illustrates, odd-index replicas of the cubic (Pm-3m) cell do not allow octahedral single-mode rotation. For example, all the octahedral rotation modes denoted by the simple-mode Glazer notation [51] such as $a^0a^0b^-$, require a cell size of at least √2x√2x2 or 2x2x2. The multi-mode DOWPs formed by more complex Glazer modes [51] are denoted, for example, as $(a^0a^0b^-)_1^2(a^0a^0b^+)_2^3(a^0a^0b^-)_3^4$ where subscript and superscript integer indices denote the $SrTiO_3$ layers along [001] direction, i.e., the first and the second $ABO_3$ layers along [001] direction have $a^0a^0b^-$ mode, while the second and the third layers have $a^0a^0b^+$ mode. Such multi-mode DOWPs can be represented by cell sizes of at least 2x2x4. Whereas none of those mentioned above, rotation modes can be realized along an odd-size axis, meaning that all possible intrinsic DOWPs related to octahedral rotation will be missed in the odd-size cell such as 3x3x3 or 5x5x5. Thus, any physical effect related to intrinsic non-thermal octahedral rotation DOWPs would be missed in a 5x5x5 representation used, e.g., in molecular dynamics simulations [53].
Once identifying the allowed symmetry of supercells needs to capture certain modes (here, even number replicas

$N$), one must perform supercell energy convergence tests with respect to $N$ within the symmetry allowed group until the structural and electronic properties are converged. In our calculations of multi-mode DOWPs polymorphous structures, the global lattice vectors of the supercells are fixed to the macroscopically observed a=b=c cubic shape (otherwise, unconstrained relaxation will converge to the low-temperature low-symmetry ground states). At the same time, all cell-internal atomic positions are allowed to relax following the calculated quantum mechanical forces. The force threshold on each atom is set to less than 0.01 eV/Å. The technical parameters are described in Supplementary section A. We use the 4x4x4 supercell consisting of 64 $ABO_3$ formula (320 atoms) units which is symmetry-compatible with all possible DOWPs we study. Relaxation is performed by applying random nudges on each atom in the 4 x 4 x 4 supercell with nudge amplitudes randomly selected between -0.15 to 0.15 Å with 0.01 Å steps. The initial lattice constant a=b=c for the polymorphous case was taken as the value obtained by minimizing the energy vs. volume of the monomorphous case. The lattice constant was subsequently allowed to relax in the polymorphous case with the energy change less than 1 meV/atom. The present approach of multi-mode DOWPs polymorphous nudging creates upon relaxation a natural distribution of locally disordered motifs. Whereas such polymorphous structures do inherit upon relaxation some of the local distortion motifs from the low-temperature low symmetry (orthorhombic, tetragonal) corner-shared phases. Unlike the latter, these local motifs in the nudged polymorphous structures are not necessarily spatially ordered. As such, we believe that these polymorphous networks provide useful "structural approximants" for describing the electronic structure of realistic cubic perovskites before thermal agitation sets in. The thermally induced disorder is then added at the DFT molecular dynamics stage (Sec. V), building upon the intrinsic polymorphous starting structure.

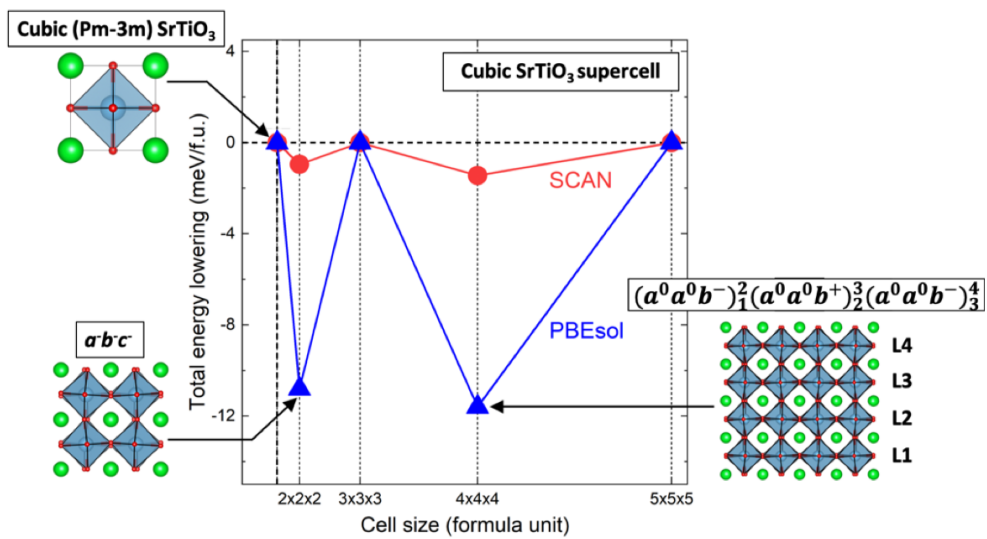

**Figure 1.** DFT internal energies of the minimal cell and enlarged 2x2x2, 3x3x3, 4x4x4, and 5x5x5 supercells of cubic $SrTiO_3$ at equilibrium lattice constant. The internal energy from the minimal cell has been chosen as the reference. Static rotation DOWPs are symmetry-forbidden in the odd-size supercells. Therefore, the minimal 1x1x1 cell, as well as 3x3x3 and 5x5x5 supercells, all show no rotation DOWPs and the unchanged energy. The 2x2x2 supercell shows energy lowering due to an $a^-b^-c^-$ (Glazer notation) octahedral rotation, while the 4x4x4 supercell shows further internal energy lowering due to $(a^0a^0b^-)_1^2(a^0a^0b^+)_2^3(a^0a^0b^-)_3^4$ multi-mode rotation. Different DFT exchange-correlation functionals give different energy lowering (SCAN [57] in red and PBEsol [58] in blue), but consistent energy-lowering trend and rotation modes among all cells studied. The equilibrium lattice constant $a_0$ = 3.907 Å is obtained from DFT using PBEsol functional and in good agreement with the experimental value [59].

## IV. Effects of intrinsic DOWPs on internal energy and local structure

### A. Identification of single-mode distortions that lower the internal energy of specific compounds

Whereas intrinsic DOWPs were defined geometrically rather early, e.g., in the classic works of Glazer and others [51,60], the question of whether they actually lower the total internal energy $U_0$ of particular cubic compounds requires compound-dependent total energy calculations. As Fig. 2 illustrates, not all the DOWPs in cubic compounds result in energy lowering with respect to undistorted, nominal cubic (Pm-3m) structure. Not all cubic oxide $ABO_3$ or halide $ABX_3$ compounds react in a similar way to specific DOWP displacements. Being the result of internal energy lowering, such DOWPs represent the expression of various bonding mechanisms such as lone-pair stereochemical effects [18,25,61,62], Jahn-Teller distortions (i.e., electronic degeneracy removal) [30], or semiclassical Jahn-Teller-like [30] effects, thus being naturally bonding-dependent.

Fig. 2 indicates that the energy lowering due to single-mode DOWPs is strongly compound-dependent: in cubic $BaTiO_3$, only the Ti off-center displacement mode can result in energy lowering with respect to undistorted structure, whereas, in cubic $SrTiO_3$, $CsPbI_3$, and $CaTiO_3$, only octahedral rotation mode DOWPs result in the internal energy lowering. The energy lowering can be significant, e.g., the rotational DOWPs can lead to more than 200 meV/f.u. energy lowering relative to the undistorted structures. These energy-lowering rotation angles with respect to the undistorted structure are 4° for the $a^-a^-a^-$ mode in $SrTiO_3$, 13° for the $a^0a^0b^-$ mode in $CsPbI_3$, and 8° for $a^-a^-a^-$ mode in $CaTiO_3$, as what is found in other perovskites compounds [26,61,63–68].

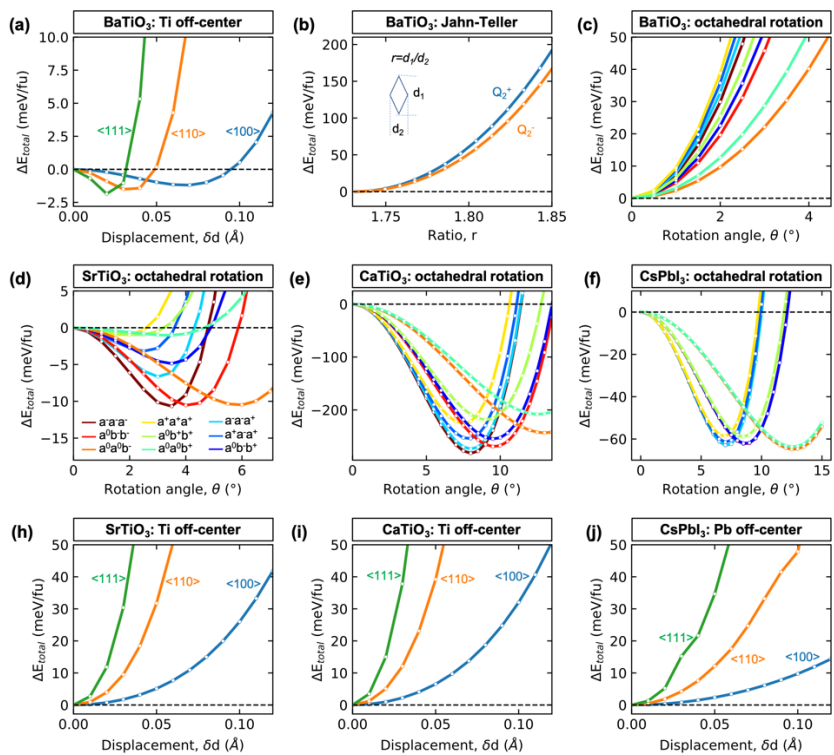

**Figure 2.** Illustrations of some of the symmetry-allowed single-mode distortions on internal energy in cubic oxide and halide perovskites. All calculations use DFT with PBEsol exchange-correlation functional. Single-mode distortion is applied one at a time with frozen-in amplitude, and the total energy is evaluated in DFT without further relaxation. The total internal energies on the y-axis is shown with respect to that of undistorted single cubic (Pm-3m) cell. (a, h-j) Energy changes as a function of B-site off-center displacements are shown in cubic $BaTiO_3$, $SrTiO_3$, $CaTiO_3$, and $CsPbI_3$. The frozen-in Ti off-center along <111> direction (Δd = 0.035 Å) results in the most significant energy lowering (1.8 meV/fu) among all the considered configurations. (b) Pseudo-Jahn-Teller distortion $Q_2^+$ and Jahn-Teller distortion $Q_2^-$ in cubic $BaTiO_3$; (c–f)

octahedral rotation modes using the simple Glazer notation [51] in cubic BaTiO$_3$, SrTiO$_3$, CaTiO$_3$, and CsPbI$_3$. Breathing mode and lattice expansions along one, two, and three directions (defined as a, ab, abc directions) are shown in Fig. S-1 and S-2 in Supplementary section A.

**B. Multi-mode deformations: internal energy lowering and the emergence of polymorphous networks**

Besides the single-mode energy lowering distortion illustrated in Fig. 2, intrinsic DOWPs can also be multi-mode DOWPs (as described in section "The type of systems discussed in the current symmetry-breaking approach: Intrinsic and dynamic") in polymorphous cubic BaTiO$_3$, SrTiO$_3$, CaTiO$_3$, CsPbI$_3$, and CsSnI$_3$. Such structures that have been cubically constrained and minimized in total internal energy show, sometimes, more considerable internal energy lowering due to multi-mode DOWPs, as indicated in Table 2.

**Table 2.** The DFT energy lowering $\Delta E_{P-M}$ due to multi-mode distortions in polymorphous structures compared with the energy of the nominal cubic (Pm-3m) structure was calculated by using the PBEsol functional. These polymorphous structures with multi-mode distortions were obtained by minimization of internal forces of the nudged, cubically shaped 4x4x4 supercells. The table also gives the Goldschmidt factor ($t_{eff}$) using Shannon ionic radii. The listed oxide perovskites have *p*-like valence band maximum (VBM) and *d*-like conduction band minimum (CBM), and the halide perovskites listed here have *sp*-like VBM and *p*-like CBM.

| Energy lowering in cubic oxide perovskites | | | Energy lowering cubic halide perovskites | | |
|---|---|---|---|---|---|
| (*p*, *d**) ABO$_3$ | $t_{eff}$ | $\Delta E_{P-M}$ (meV/f.u.) | (*sp*, *p**) ABX$_3$ | $t_{eff}$ | $\Delta E_{P-M}$ (meV/f.u.) |
| CaTiO$_3$ | 0.97 | -360 | CsSnI$_3$ | 0.85 | -38 |
| SrTiO$_3$ | 1.00 | -12 | CsPbI$_3$ | 0.85 | -69 |
| BaTiO$_3$ | 1.06 | -7 | KPbI$_3$ | 0.80 | -280 |
| SrZrO$_3$ | 0.95 | -21 | RbPbI$_3$ | 0.82 | -180 |
| PbZrO$_3$ | 0.96 | -44 | RbGeI$_3$ | 0.95 | -45 |
| PbTiO$_3$ | 1.02 | -6 | CsGeBr$_3$ | 1.01 | -40 |
| / | / | / | CsGeI$_3$ | 0.98 | -30 |
| / | / | / | CsSnBr$_3$ | 0.86 | -10 |
| / | / | / | CsPbF$_3$ | 0.91 | -50 |
| / | / | / | CsPbBr$_3$ | 0.82 | -65 |

Table 2 also gives results for cubic ABO$_3$ oxide having *p*-like valence band maximum (VBM) and *d*-like conduction band minimum (CBM), as well as halide perovskites ABX$_3$, showing significant energy lowering due to multi-mode distortion within supercells. We anticipate that the multi-mode DOWPs can exist in other cubic perovskites with different band-edge states, such as LaVO$_3$ with (*d*, *d**) states [32]. In the following sections, we mainly discuss the DOWPs and the corresponding effect on band gap by taking ATiO$_3$ (A$^+$= Ca, Sr, Ba) and CsBI$_3$ (B$^{2+}$= Pb, Sn) as examples. The multi-mode distortions can lower the energy more than the single-mode distortions because of a few reasons. First, instead of having a single-mode with a large, uniform tilting angle and/or displacement throughout the lattice, it is sometimes energetically advantageous to have a few modes covering a range of tilting angles or displacements. Given the similar energy lowering for a single-mode distortion with a large tilting angle and a few modes with small tilting angles, we anticipate that modes such as a$^-$a$^-$a$^-$, a$^0$a$^0$b$^-$ and a$^0$b$^-$b$^-$ modes in SrTiO$_3$, a$^0$a$^0$b$^-$ and a$^0$a$^0$b$^+$ modes in CsPbI$_3$, as well as a$^-$a$^-$a$^-$, a$^0$b$^-$b$^-$, and a$^-$a$^-$a$^+$ modes in CaTiO$_3$ would coexist. The similar energy lowering of different single-mode distortions might lead to a mixture of different tilting modes with energy lowering. Indeed, as depicted in Fig. 1, the mixture mode (a$^0$a$^0$b$^-$)$_1^2$(a$^0$a$^0$b$^+$)$_2^3$(a$^0$a$^0$b$^-$)$_3^4$ lower energy (by 1 meV/f.u.) than the 2 x 2 x 2 supercell with the single a$^-$b$^-$c$^-$ mode. Second, a larger supercell can enable symmetry-breaking modes absent in a smaller cell that has restricted single-mode distortion. Again, SrTiO$_3$ supercell has (a$^0$a$^0$b$^-$)$_1^2$(a$^0$a$^0$b$^+$)$_2^3$(a$^0$a$^0$b$^-$)$_3^4$ multi-mode DOWPs not allowed in 2 x 2 x 2 cell (Fig. 1). The x, y, and z tilting angles in the 4 x 4 x 4 supercell are 3.5°, 5.9°, and 4.9°, respectively, close to the values in the relaxed 2 x 2 x 2 supercell with $a^-b^-c^-$ single-mode distortion (Fig. 1).

### C. Distribution of multi-mode tilting and displacements in polymorphous structures

Fig. 3 shows statistics of the multi-mode distortions in polymorphous oxide perovskites $BaTiO_3$, $SrTiO_3$, and $CaTiO_3$, while Fig. 4 provides equivalent results for polymorphous structures of halide perovskites $CsPbI_3$ and $CsSnI_3$. Here, we focus on the B-site off-center displacements and the octahedral tilting angles with respect to <100>, <010>, and <001> vectors, revealing distributed distortions of local motifs. In Fig. 3 and Fig. 4, the depicted distributions of octahedral tilting are different from the illustration shown in Ref. [28], which only shows the distribution of all tilting angles without differentiating the distribution along each direction. Here, by showing the corresponding distributed octahedral tilting angle with respect to <100>, <010>, and <001> vectors (defined as x, y, and z directions), it can be seen if the DOWPs are single-mode or multi-mode. The results can be summarized as follows:

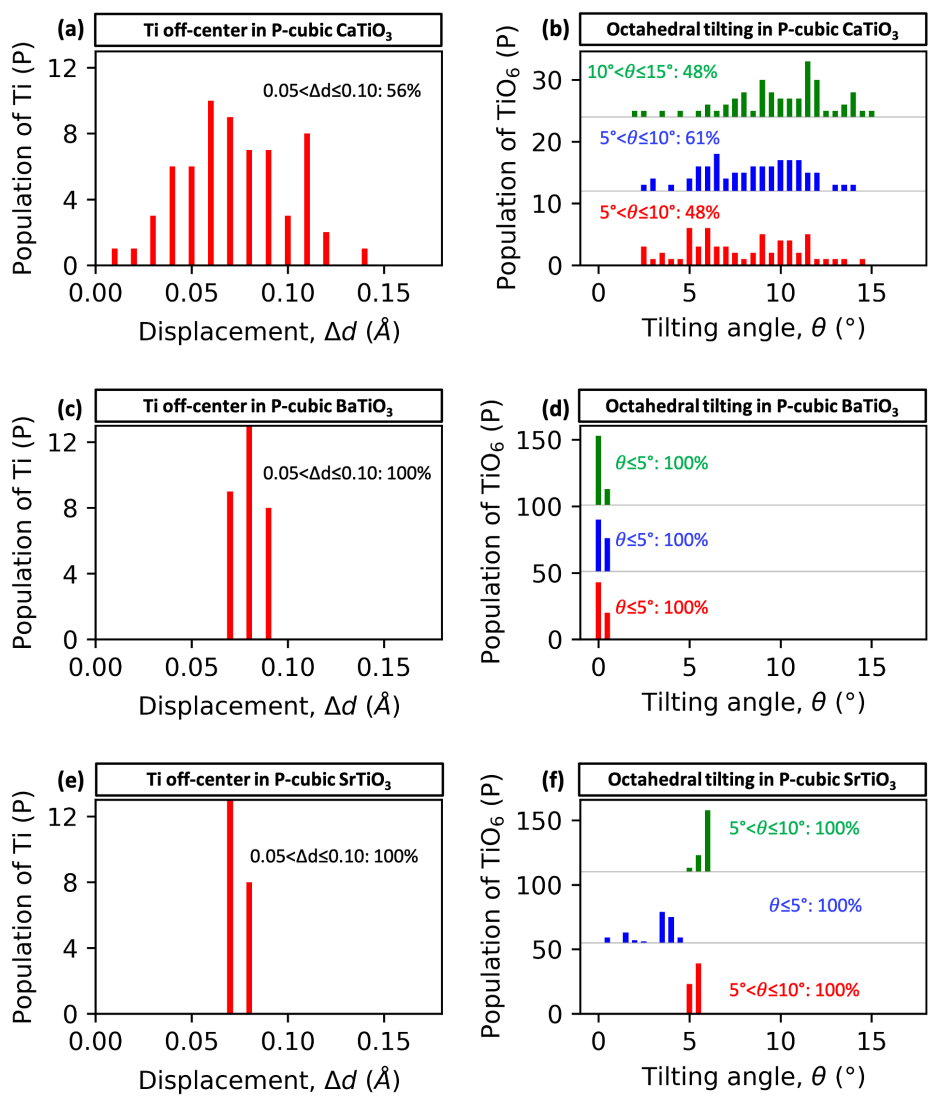

**Figure 3.** Statistical distributions of intrinsic multi-mode DOWPs in oxide perovskites. Shown are the number of Ti atoms with Ti off-center of different amplitudes (a, c and e) and octahedra with varying angles of tilting (b, d and f) with respect to Cartesian axes x (red bars), y (blue bars), and z (green bars) in 4x4x4 polymorphous cubic (P-cubic) $CaTiO_3$, $BaTiO_3$ and $SrTiO_3$ supercells. The major range of distribution of off-center and octahedral tilting angles and their percentages were depicted in panels.

i. CaTiO$_3$ (Fig. 3a and b): The Ti displacements and octahedral tilting angles are distributed over relatively large ranges, though the major Ti displacements and octahedral tilting angles occur up to ~0.15 Å and ~15°, respectively. These Ti displacements and octahedral tilting distributions indicate the multi-mode DOWPs in the polymorphous network.

ii. BaTiO$_3$ (Fig. 3c and d): Ti displacements are mostly along {111} directions (close to the values of Ti displacements in the low-temperature orthorhombic structure [69]). Furthermore, there is negligible octahedral tilting, indicating a single ferroelectric intrinsic DOWPs mode (Fig. 2). Indeed, experimental PDF [70] suggests that this locally rhombohedral-like structure with Ti ferroelectric along {111} direction exists in paraelectric cubic BaTiO$_3$.

iii. SrTiO$_3$ (Fig. 3e and 3f): The Ti displacements and the octahedral tilting angles are distributed over narrow ranges (Ti displacements up to ~0.07 Å, tingle angle in a range of 0–60 along x, y, z axes). Note that the Sr atoms also slightly displace (up to 0.04 Å) from the perfect Wycoff position within a narrow range (Table 1).

iv. CsSnI$_3$ and CsPbI$_3$ (Fig. 4): The Sn or Pb displacements and octahedral tilting angles are distributed over a wide range (up to ~0.2 Å). Note that the B-site displacements and octahedral tilting angles in cubic CsPbI$_3$ are relatively larger than those in cubic CsSnI$_3$.

In short, there is a distribution of local motifs (here, B displacements and octahedral tilting angles) in cubic oxide and halide perovskites, forming a polymorphous network. The smaller Goldschmidt factor (<1.0) implies larger octahedral tilting in a cubic structure, such as CaTiO$_3$.

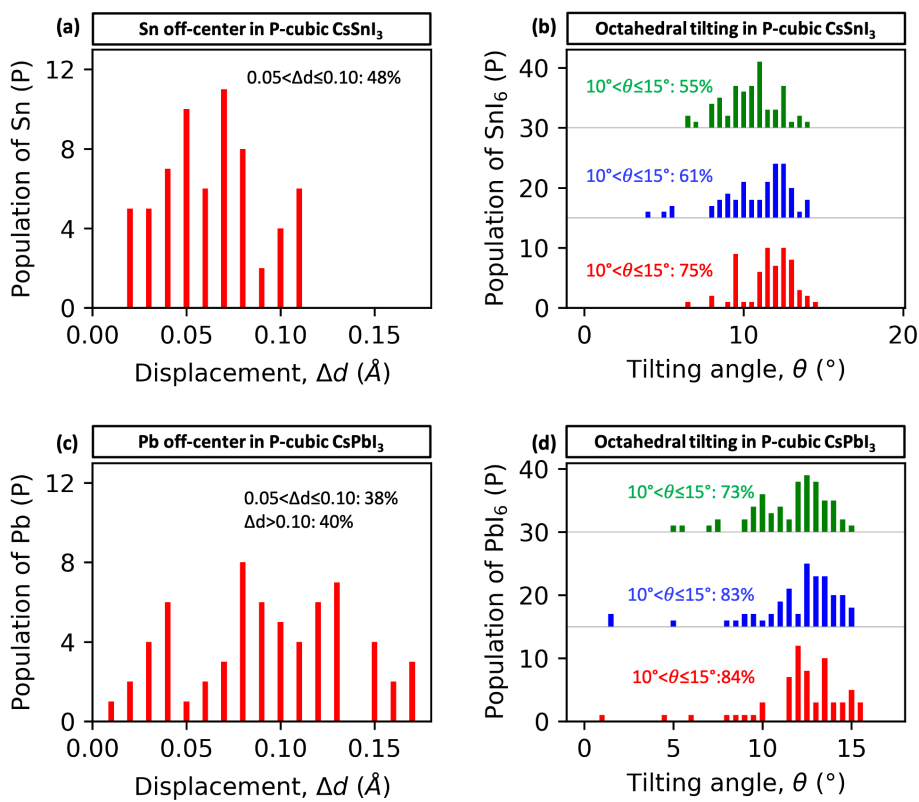

**Figure 4.** Statistical distribution of intrinsic multi-mode DOWPs in halide perovskites. Shown are the number of Sn/Pb atoms with Sn/Pb off-center of different amplitudes (a and c) and octahedra with varying angles of tilting (b and d) with respect to Cartesian axes x (red bars), y (blue bars), and z (green bars) in 4x4x4 polymorphous cubic (P-cubic) CsSnI$_3$ and CsPbI$_3$ supercells. The major range of distribution of off-center and octahedral tilting angles and their percentages were depicted in panels.

## D. Comparison of pair distribution function based on polymorphous cubic SrTiO$_3$ with multi-mode distortions to experimental observation

To examine whether the multi-mode DOWPs are physically reasonable, we compare in Fig. 5 the calculated XRD diffraction and PDF obtained from the polymorphous cubic SrTiO$_3$ with those experimentally measured [71] at 225 K (calculation details in Supplementary section A). The width of the calculated XRD peaks conceals information on the statistical distribution of interatomic distances and short-range order. This information, not addressed in the present paper, awaits to be mined in the form of SRO parameters and PDF data as reported in Ref. [28]. The PDF is a local probe that reflects the distribution of atom-atom distances, thus a good way to determine the local atomic configurations and distortion of local motifs within the system. As already noted, fitting the diffraction data in the Rietveld process to a single formula cubic (Pm-3m) cell [72] can miss the existence of intrinsic DOWPs that evidently lead to energy lowering (Fig. 1) when it is allowed to occur. We see in Fig. 5 that the DFT predicted PDF for SrTiO$_3$ shows good agreement with the experiment. There are excellent agreements for halide perovskites between the calculated PDF from T = 0 DFT and the measured PDF at finite moderate temperatures [73] such as MAPbI$_3$, etc. Ref. [28] provides additional evidence to the reasonableness of the DOWP structures deduced from DFT. We note, however, that the tilting in SrTiO$_3$ is relatively small, as suggested by the tolerance factor being close to 1, so the dependence of PDF on the DOWPs is not as critical as in halide perovskites [28] or FeSe [27].

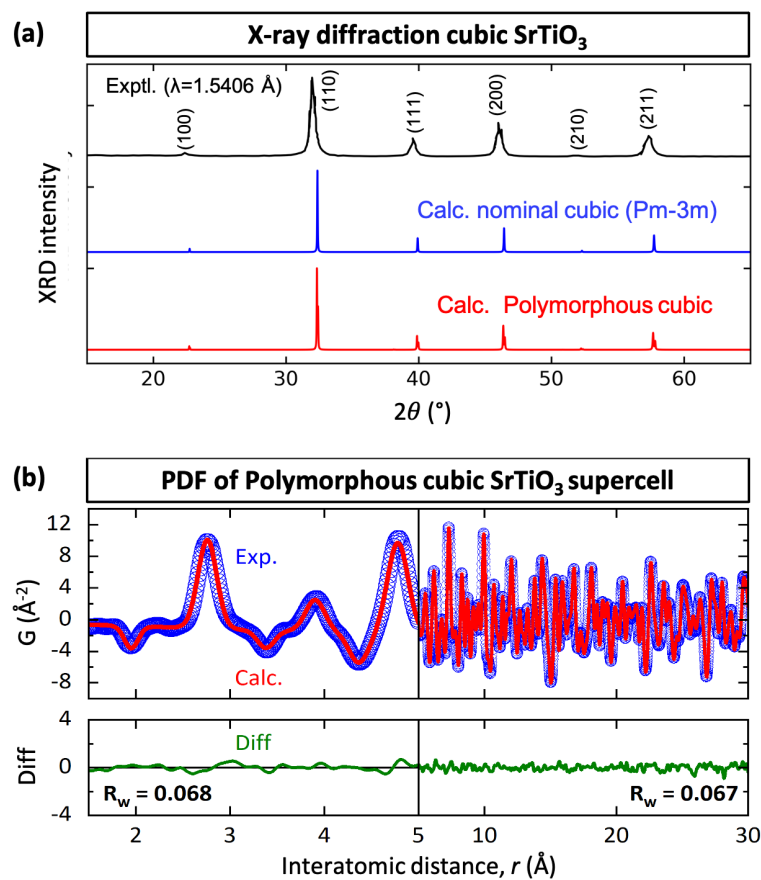

**Figure 5.** (a) Comparison of X-ray diffraction (XRD) patterns among experimentally measured [74] SrTiO$_3$ single crystal (black line), simulated XRD based on the calculated nominal cubic SrTiO$_3$ (blue line) and polymorphous 4x4x4 supercell with multi-mode DOWPs by using SCAN functional. The XRD peaks in the polymorphous network are slightly broadened relative to the monomorphous structure due to the absence of a strong correlation between the local motifs. (b) Comparison between experimentally measured [71] pair distribution function (PDF) of cubic SrTiO$_3$ at 225 K (empty blue circles) and DFT calculated PDF at T = 0 (red lines) based on polymorphous 4x4x4 supercell optimized by SCAN functional.

The overall R-factors are given for interatomic distance 0–5 Å (left) and 5–30 Å (right), revealing the derivation of the predicted PDF from the experimental measured PDF.

## V. Effects of intrinsic single- and multi-mode DOWPs on band gaps

### A. Turnover from blueshift at small single-mode deformation to redshift at large single-mode deformation

Using the structures with different single-mode distortions (viz. Fig. 2) within frozen-in amplitudes in cubic 2 x 2 x 2 supercells, we have calculated non-perturbatively the corresponding band structures. Here, we deviate from the traditional perturbative treatment where electron-phonon interactions are summed explicitly (often with specific truncations). Figs. 6 and 7 show for a few compounds the band gap shifts due to specific deformation modes as a function of the amplitudes of distortion. Fig. 6(a–c) shows that the octahedral rotation in oxide perovskites leads to redshift when the tilting angle is larger than ~8°. In contrast, smaller rotation angles and other modes (e.g., B-atom off-center displacements and Jahn-Teller distortion) lead to band gap blueshift. For the halide perovskites (Fig. 7a and b), octahedral rotation between 0 and 15° always leads to band gap blueshift relative to the nominal cubic (Pm-3m) structure. For instance, for cubic $SrTiO_3$, the 15°-rotation angle for $a^0b^+b^+$ mode leads to a DFT gap of 1.57 eV (Fig. 6c), presenting 0.13 eV blueshift relative to the untilted, ideal structure. The $a^0b^+b^+$ single-mode distortion in cubic $CaTiO_3$ with 15°-rotation angle leads to a DFT band gap of 0.93 eV (Fig. 6b), indicating a 0.75 eV redshift compared to the undistorted cubic (Pm-3m) structure. Table 1 provides a succinct summary. These trends are explained next.

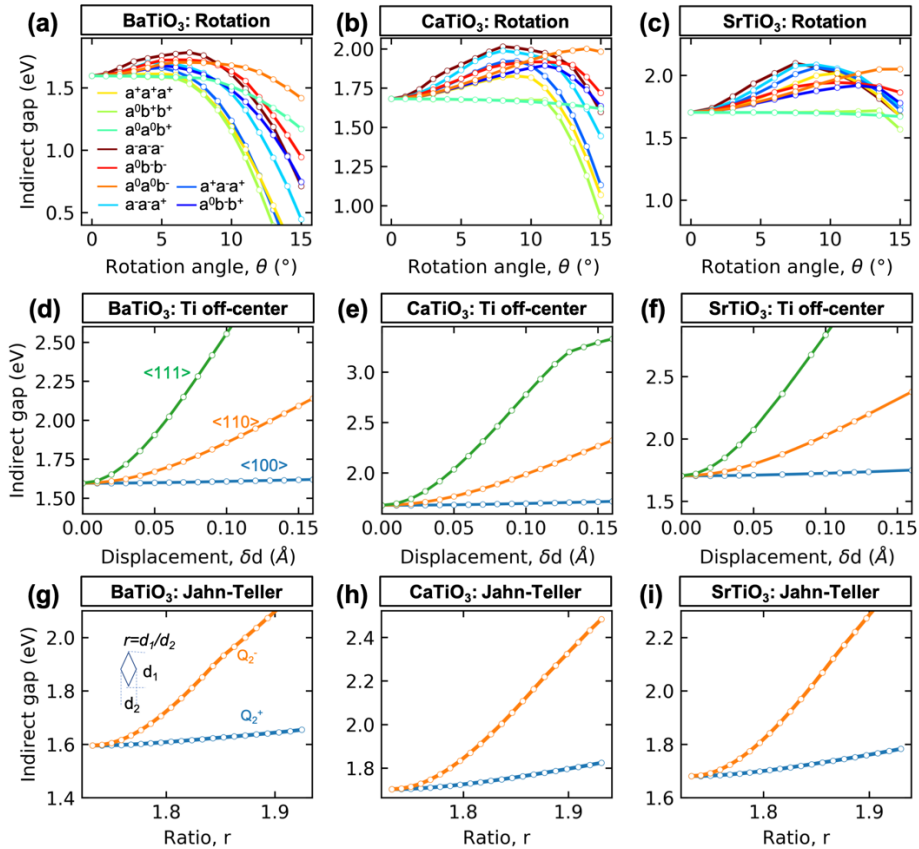

**Figure 6.** Cubic oxide perovskites. DFT band gap changes as a function of different single-mode distortions with frozen-in amplitudes in cubic 2x2x2 supercells of $BaTiO_3$, $CaTiO_3$, and $SrTiO_3$: (a-c) illustrate the gap shift by octahedral rotation mode, whereas (d-f) illustrate the gap shift by B-site off-center displacement along <100>, <110>, and <111> directions. Panels (g-i) illustrate the effect of Jahn-Teller modes on the band gaps. Breathing modes and lattice expansion modes are also described in Supplementary Figure S-3. The legends for the off-center displacements, rotation, and the Jahn-Teller modes are depicted in inserts (a), (d), and (g), respectively.

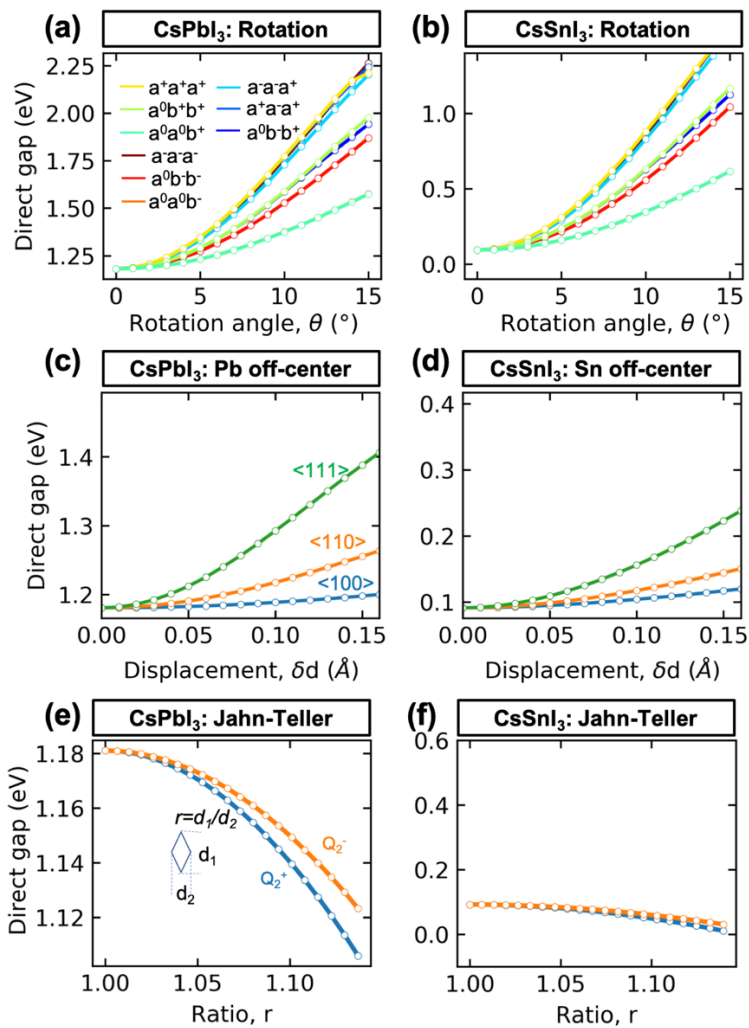

**Figure 7**. Cubic halide perovskites. DFT band gap as the function of different single-mode distortions with frozen-in amplitudes in cubic 2x2x2 supercells of $CsPbI_3$ and $CsSnI_3$: (a and b) illustrate the effect of octahedral rotation modes on the band gap, (c and d) depict the band gap shift due to B site off-center along <100>, <110>, and <111> directions, whereas (e and f) show the effect of Jahn-Teller modes on the band gaps of cubic $CsPbI_3$ and $CsSnI_3$. Breathing modes and lattice expansion modes are illustrated in Supplementary Fig. S-4.

**B. Blueshifts of the indirect gap and redshifts of the direct gap in cubic oxide $CaTiO_3$ as results of single-mode octahedral rotation**

To understand the aforementioned band gap blueshift or redshift in oxide and halide perovskites due to the intrinsic DOWPs, we take cubic $CaTiO_3$ and $CsPbI_3$ as examples. Fig. 8 shows the effective band structures (EBS) [75] and the energy diagram before and after octahedral rotation for the most energetically favorable frozen single Glazer mode. The EBS showed in Fig. 8a and b are unfolded band structures transforming rigorously the band structure of large real space supercell to the Brillouin Zone appropriate to the primitive real space cell (details in Supplementary section A). As Fig. 8a illustrates, the octahedral rotation angle of $a^-a^-a^-$ mode at 5° in $CaTiO_3$ leads to an increase in the indirect band gap at R-Γ and a decrease in the direct band gap at Γ-Γ compared with values in the untilted structure. These shifts reveal a transition from the indirect band gap in the untilted structure to a nearly degenerated direct-indirect band gap. For the cubic halide CsPbI$_3$, the rotation angle of a$^0$a$^0$b$^-$ mode at 5° results in an increase in the direct band gap at R–R with respect to undeformed structure.

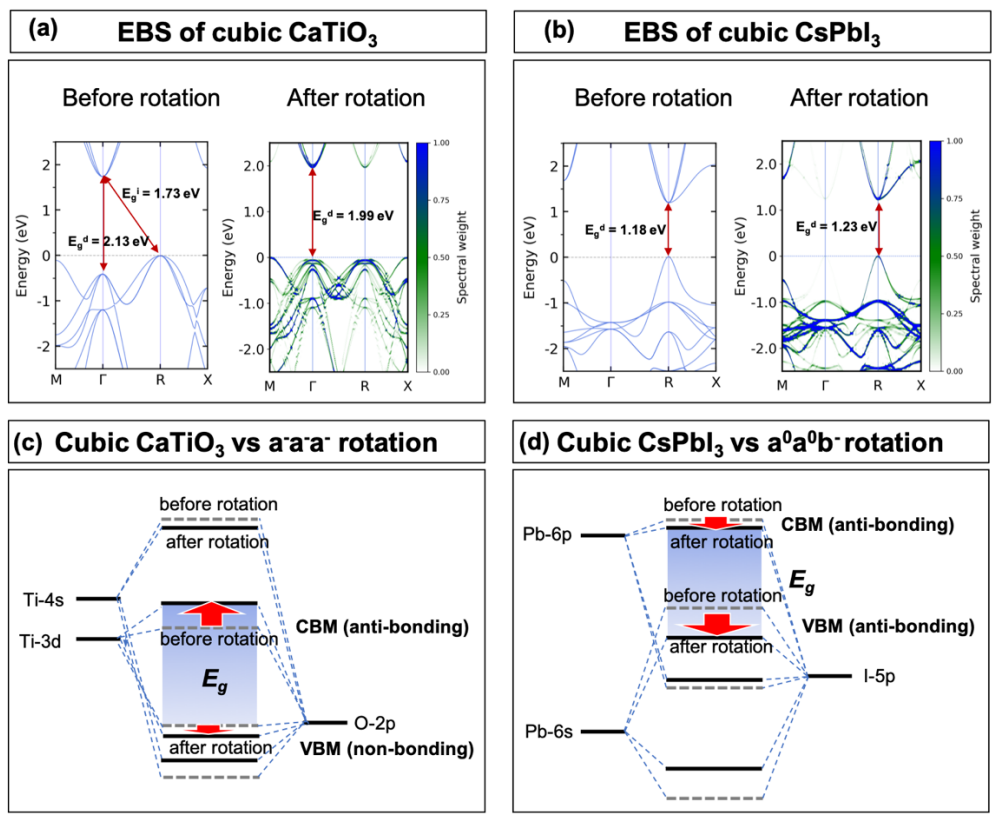

**Figure 8.** (a, b) The unfolded effective band structure (EBS) from 2x2x2 supercell to primitive cell (1fu/cell) of cubic CaTiO$_3$ and CsPbI$_3$. Illustration of energy diagram of (c) cubic CaTiO$_3$ with a$^-$a$^-$a$^-$ energy lowering rotation single-mode and (d) cubic CsPbI$_3$ with a$^0$a$^0$b$^-$ energy lowering rotation mode. The amplitude of rotation angle in both (a) and (b) is frozen at 5°. The calculated indirect band gap ($E_g^i$) and direct band gap ($E_g^d$) are depicted in (a and b). The horizontal dash lines at zero in (a and b) band structures refer to the DFT calculated highest occupied level.

Fig. 8c and d show the energy level diagrams for cubic CaTiO$_3$ and CsPbI$_3$ (referenced to the core-level of Ti-1$s$ and Pb-1$s$) before and after octahedral rotation, which explain the above-noted DFT results. From Fig. 8c, we see that for the cubic oxide CaTiO$_3$, the VBM is a non-bonding state (O-2$p$), and the CBM is an anti-bonding state (majority Ti-3$d$), whereas, for the cubic halide CsPbI$_3$, both VBM and CBM are anti-bonding states (Pb-6$s$ – I-5$p$ at VBM and Pb-6$p$ – I-5$p$ at CBM as illustrated in Fig. 8d). We find that the octahedral rotation in CaTiO$_3$ leads to strong orbital hybridization between Ti-3$d$ and O-2$p$ states, which pushes anti-bonding states at CBM into higher energy level comparing to CBM level in the undistorted structure. In contrast, the VBM of distorted CaTiO$_3$ downshifts into slightly deeper energies than the VBM level in undistorted structure due to the weakened repulsion of Ti-4$s$ and O-2$p$ non-bonding state. Besides, the maximum occupied valance band at the Γ point upshifts because of the enhanced repulsion between Ti-3$p$ and O-2$p$ non-bonding state at Γ point. Therefore, in cubic oxide CaTiO$_3$, the R–Γ indirect gap blueshifts and Γ–Γ direct band gap redshifts as a result of octahedral rotation. While the octahedral rotation in the cubic halide enlengthens Pb–I bond length hence weakens the Pb–I coupling. The weakened coupling results in a downshift of VBM and CBM energy levels. However, VBM downshifts more than CBM because VBM is more sensitive to the enlengthened bond length due to inner-shell Pb-6$s$ states contributed to the VBM. The relatively large downshift of VBM leads to the R–R band gap blueshift in cubic halide CsPbI$_3$ [76,77]. The same mechanism persists in a polymorphous network with multi-mode distortions.

## C. Emergence of a direct band gap component in CaTiO$_3$ and SrTiO$_3$ due to single-mode octahedral rotation

We have seen that in cubic oxide CaTiO$_3$, there is a transition from indirect to nearly direct band gap due to the octahedral rotation for a$^-$a$^-$a$^-$ single Glazer mode (Fig. 8a). Considering the unfolded band structure (Fig. 9a), we also find that other octahedral rotation modes, e.g., a$^0$a$^0$b$^-$, a$^+$a$^-$a$^+$, and a$^-$a$^-$a$^+$, in cubic CaTiO$_3$ lead to such indirect-to-direct band gap transition. Consequently, as can be seen from Fig. 8a and Fig. 9a, besides the already existing minimal R-Γ indirect gap, intrinsic DOWPs lead to minimal Γ-Γ direct band gap component as a secondary valley. Note that such a direct-in-momentum gap component is not a consequence of the folding mechanism in the supercell, because it survives in the EBS. A similar phenomenon is seen in cubic SrTiO$_3$.

To analyze the relationship between the direct gap component and the rotation DOWPs, Fig. 9b shows several cubic SrTiO$_3$ with frozen-in a$^0$b$^-$b$^-$ mode with a rotation angle of 0°, 2.5°, and 5°. The increase in rotation amplitude causes the VBM spectral intensity at Γ to increase, but VBM spectral intensity at R diminishes. These trends signal the buildup of a significant direct band gap component in cubic SrTiO$_3$ due to tilting DOWPs.

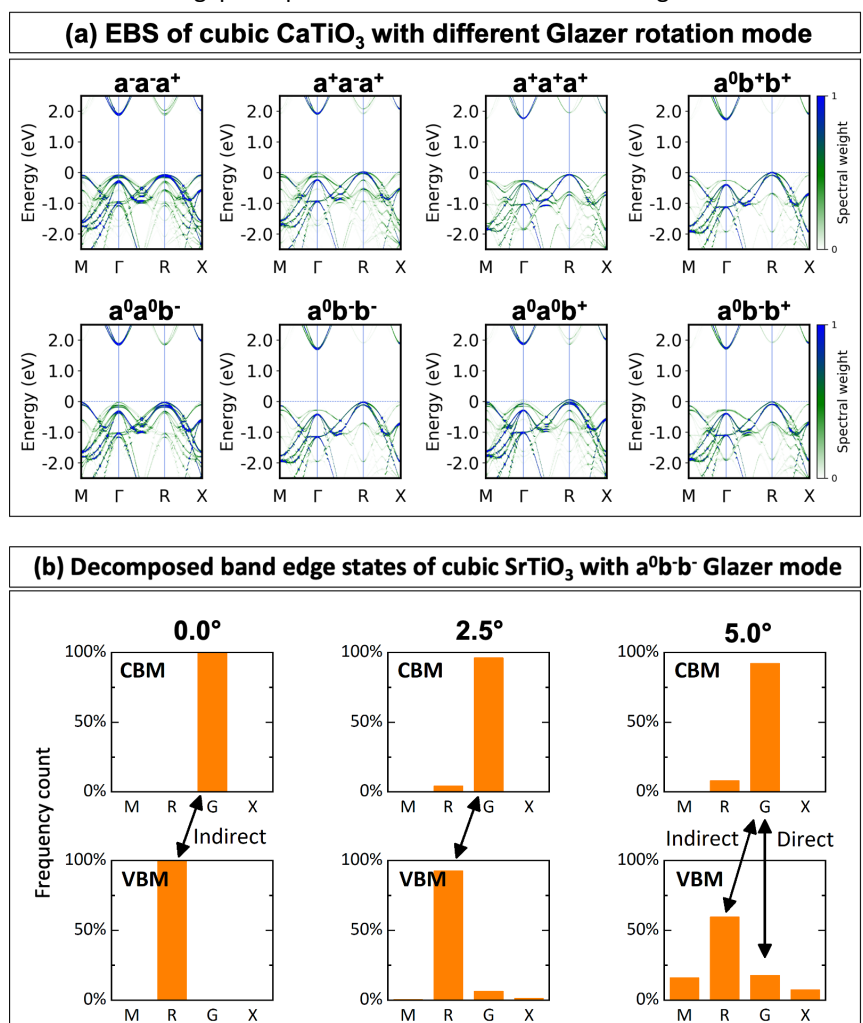

**Figure 9.** The effective band structure (EBS) of (a) CaTiO$_3$ in 2x2x2 supercell with frozen-in single Glazer modes with frozen-in amplitude of rotation angle 5°. CaTiO$_3$ has an indirect-to-direct band gap transition for the a$^-$a$^-$a$^+$ mode, while other Glazer modes result in an indirect band gap from the VBM at the R point to the CBM at the Γ point. (b) Evolution of band edge states at high symmetry K-points in cubic SrTiO$_3$ as a function of rotation angle. The y-axis shows the percentage of the counted spectral weights at different high symmetry K-points. The VBM of cubic SrTiO$_3$ has more and more Γ components as rotation angle increases, meaning a larger and larger direct gap component.

**D. Effects of single-mode and multi-mode non-thermal deformations on band gaps**

Fig. 10 summarizes for several cubic ABX$_3$ compounds (i) the minimal band gap from monomorphous (Pm-3m) cell (shown in blue), (ii) the gap resulting from allowing single-mode DOWPs that lead to the largest absolute energy lowering (shown in green), and (iii) the band gap due to multi-mode distortion in the polymorphous network (in red). The lowest energy single-mode distortions in cubic CaTiO$_3$, BaTiO$_3$, CsSnI$_3$, and CsPbI$_3$ are a$^-$a$^-$a$^-$, Ti off-center along <111> direction, a$^0$a$^0$b$^-$, and a$^0$a$^0$b$^-$, respectively. As shown in Fig. 10, the cubic (Pm-3m) structures without symmetry breaking have the smallest band gaps. The blueshift of these band gap values due to single-mode distortions are in the range of 0.05 eV for BaTiO$_3$ to 0.39 eV for CaTiO$_3$ with respect to their corresponding cubic (Pm-3m) structure without symmetry breaking. The cubic structures with multi-mode DOWPs in polymorphous network show much larger blueshift values from 0.11 eV for BaTiO$_3$ to 0.55 eV for CsSnI$_3$. Local symmetry breaking, either single-mode or multi-mode DOWPs, can lead to significant band gap blueshift.

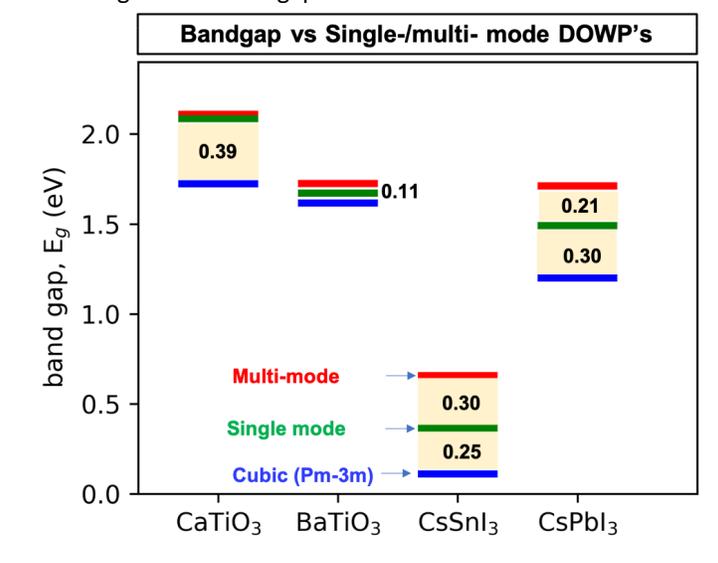

**Figure 10.** Minimal band gap E$_g$ for (i) cubic Pm-3m (blue line) structure without distortion, (ii) the cubic structure with frozen in single-mode distortions that results in the lowest energy lowering (green line), and (iii) multi-mode DOWPs in the polymorphous network (red bar) of CaTiO$_3$, BaTiO$_3$, CsSnI$_3$, and CsPbI$_3$. The shaded regions highlight the band gap shift. The gap in the cubic (Pm-3m) structures refers to the indirect transition for oxide perovskites (R-Γ) to the (R-R) direct band gap for halide perovskites.

## VI. Effect of temperature-induced distortion in molecular dynamics on band gaps

**A. DFT molecular dynamics: Temperature- enhanced deformations in CaTiO$_3$ and SrTiO$_3$**

The free energy of the vibrating lattice F = U$_0$ - TS at finite temperature can be studied using MD simulations. As the temperature increases, the free energy decreases due to the –TS term, which stabilizes the local distortion at finite temperature. In ab initio MD (AIMD) methodology, the finite-temperature dynamical trajectories are generated using forces obtained directly from electronic structure calculations. Therefore, phonon–phonon interactions are taken into account as the response of the electronic structure to deformations (represented in textbooks by the electron-phonon interactions). For several different temperatures, we extracted for each temperature $N$ snapshots (N = 2000 for CaTiO$_3$ and N = 2471 for SrTiO$_3$) from the trajectory between ~1 ps - 2 ps in AIMD calculations for cubic CaTiO$_3$ in a 4x4x4 supercell and cubic SrTiO$_3$ in a 5x5x5 supercell (from Ref. [53]). Note that in a 5x5x5 supercell, the AIMD simulation of Ref [53] will not account for non-thermal octahedral tilting symmetry breaking, as illustrated in Fig. 1 but can allow for Sr and Ti off-center displacements. In contrast, the non-zero tilting angles seen in AIMD simulations might originate from the thermal motions of oxygen atoms. Thus, the restriction to 5x5x5 supercell permits thermal but excluded non-thermal octahedral tilting, whereas 4 x 4 x 4 cell allows both.

Since the distortions apparent from MD are rather complex and difficult to be identified as specific, pure symmetry modes, we performed instead for each snapshot a statistical analysis of effective local octahedral tilting angles with respect to x, y, and z axes of the cubic cell as shown in Fig. 3 and Fig. 4. A similar analysis was done for Ti displacement with respect to the center of octahedra. The population (P) of tilted octahedra with specific tilting angle and Ti displacement at different temperatures, were depicted in Fig. 11.

Fig. 11 shows the thermally induced octahedral tilting and B-site displacement for cubic CaTiO$_3$ and SrTiO$_3$ at different temperatures. Significantly, the main peaks in Fig. 11a and c show a broadening of the octahedral tilting angles around non-zero tilting angle values from low to high temperatures in cubic CaTiO$_3$ and SrTiO$_3$, as well as the Ti displacements. Notably, the amplitudes of A sites (Sr or Ca) displacements are significantly enlarged at high temperatures compared to B-site displacements (Table 1). For example, the major Sr displacements start ~0.03 Å at T = 0 (polymorphous structure) to ~0.23 Å at 1400 K. The presence of snapshot averaged non-zero distortions at low temperatures represent the intrinsic DOWPs of Ti displacements. The thermal-induced octahedral tilting and Ti off-center displacements are enhanced as temperature increases. The thermally enhanced atomic displacements at finite temperatures are consistent with enlarged Li displacements (rather than diminished displacements) as temperature increase in paraelectric LiNbO$_3$ [39].

**B. The effects on band gap shift due to thermal distortion in oxide perovskites**

The enhanced thermal motions revealed by the AIMD simulations result in significant band gap changes. For instance, the averaged band gap (expressed as $\sum_{i=1}^{N} E_g(S_i)/N$ of individual configuration $S_i$ along the thermodynamic equilibrium trajectory, where the E$_g$($S_i$) refers to the allowed and forbidden band gap of $S_i$) for cubic CaTiO$_3$ and SrTiO$_3$ are shown in Table 1. The band gap redshift due to thermal motions at 2000 K in cubic CaTiO$_3$ is 1.15 eV with respect to polymorphous structure, and 0.77 eV with respect to the nominal cubic structure. Similarly, in cubic SrTiO$_3$, the corresponding band gap redshifts are 0.72 eV and 0.49 eV, respectively, at 1400 K. Evidently, thermal motions in cubic CaTiO$_3$ and SrTiO$_3$ lead to significantly band gap redshift at high temperature (>T$_c$) compared with band gap at T = 0 (monomorphous structure).

Table 1 provides both a preview and a succinct summary of the main differences between the effects of intrinsic non-thermal vs. dynamic thermal displacements on band gaps for oxide perovskites.

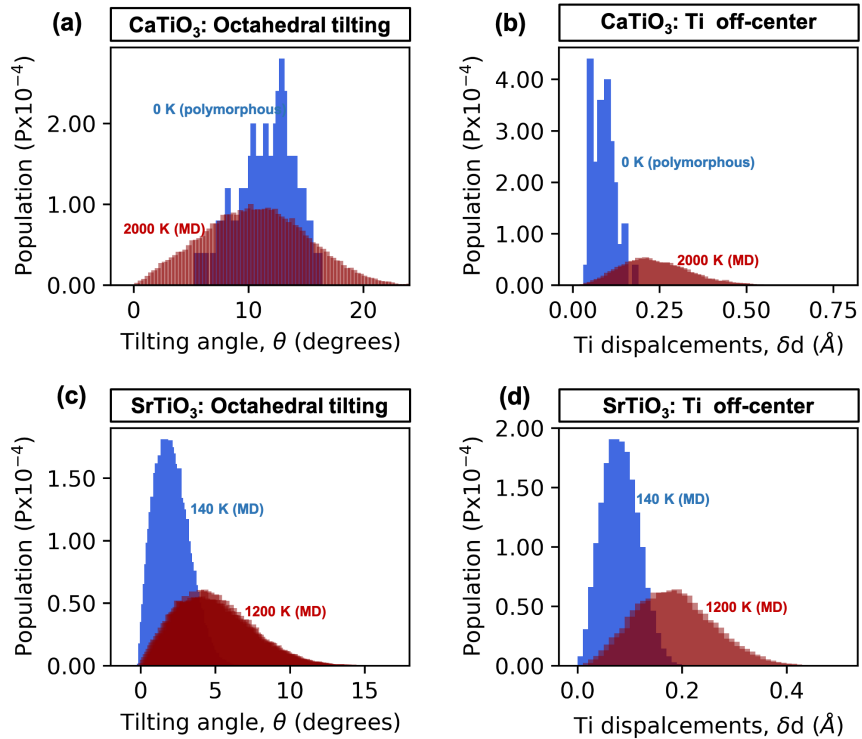

**Figure 11.** The distribution of populations of local distortions obtained from snapshots using DFT molecular dynamic simulation in NVT ensemble for cubic CaTiO$_3$ in a 4x4x4 supercell (a, b) at 2000 K (dark-red bars) using the PBEsol functional and Gamma-only k-grid, and for SrTiO$_3$ in a 5x5x5 supercell at 140 K from Ref. [53] (blue bars) and 1200 K (dark-red bars) employing the PBE functional with Tkatchenko-Scheffler dispersion correction [78] (c, d). The local distortions are octahedral tilting angle (a and c) and Ti off-center displacements (b and d). For comparison, the statistical populations multiplied by 2000 of local distortions in a 4x4x4 supercell CaTiO$_3$ polymorphous network were also depicted (a–d, blue bars). The data of SrTiO$_3$ are extracted from reference [53].

*For oxide perovskites:* Different from nominal cubic SrTiO$_3$ and CaTiO$_3$ with *p*-like VBM and *d*-like CBM, the "correlated" Mott insulators having *d*-like VBM and *d*-like CBM (*d*, *d**) in the nominal cubic structure are often high energy false metals because of the absence of DOWPs. Using the nominal cubic structure led the d-electron oxide literatures to postulate the need for dynamic electron correlation for opening the gap [32,79]. Allowing the formation of polymorphous cubic structure lowers the energy and correctly becomes an insulator even for (*d*, *d**) because of structural and electronic symmetry breaking by energy lowering, such as LaTiO$_3$ and LaVO$_3$ [30,32]. The reason is that the gap increases with tilting in the range of 0°–8° tilting angles (see Fig. 6). Finally, at really high T cubic phases, we have larger amplitudes of DOWPs for some of the Glazer modes. Such high tilting exists in MD for oxides, and reverses the direction of gap changes from blueshifts at low amplitude to redshift at high amplitudes. Thus, at such temperatures system can convert to metal.

*For halides perovskites:* These are never (*d*, *d**) but rather (*sp*, *p**) compounds, and as such, they have finite gaps even in the nominal cubic structure, albeit rather underestimated. Allowing for DOWPs lowers the internal energy and increases the band gaps. Such polymorphous structures have relatively larger band gap values, sometimes comparable to those of the low-temperature low-symmetry structures. The reason is that at low tilting angles, the gap increases with tilting. At really high T, gaps continue to rise because tilting angles increase, and the band gap is always blueshift.

## VII. Conclusions

We offer a structural approximant to the cubic perovskites that replaces the nominally cubic (Pm-3m) anzats that restricts deformations. The polymorphous ansatz is based on nudging the internal positions of a cubically shaped supercell that is initially made from replicas of the cubic (Pm-3m) monomorphous primitive cell. This results in the formation of local distortions that can be characterized as short-range order without long-range order. Such nudged cubic supercell excites multi-mode DOWPs results in lower total energy and many local motifs even before the thermal effects set in. The local intrinsic DOWPs are an expression of the nature of chemical bonding in the compounds, projecting onto many symmetry-breaking modes such as A and B site off-center, octahedral tilting/rotation, and breathing mode. The intrinsic DOWPs not only agree with local probe (such as PDF) observations in the experiment, but also show significant contribution to the changes of electronic structure. We can summarize the conclusions as follows:

(i) Comparison between halide and oxide perovskites: We learned that oxides and halides respond differently to intrinsic distortions. Cubic halide and oxide perovskites adopt different energy lowering symmetry-breaking modes. In cubic $CaTiO_3$ and $SrTiO_3$, such intrinsic DOWPs lead to a predicted yet not reported Γ-Γ direct band gap component, as a secondary valley minimum to the well-known indirect R-Γ gap. In contrast, intrinsic DOWPs only enlarge the direct band gap blueshift in halide perovskites. This finding might indicate the possibility to tune the indirect- to-direct transition in oxide perovskites by controlling a specific mode. However, in cubic halide perovskites, intrinsic DOWPs do not lead to band-edge transformation but direct band gap blue-shift. Besides, different from the flat valence band-edge states in oxide perovskites with DOWPs, the valence band-edge states of halide perovskites contributed by the lone pair states remain sharp, indicating the weak coupling between conductivity and distortion. This finding for halide perovskites provides a good principle to design optoelectrical materials with excellent conductivity.

(ii) Comparison of intrinsic vs. dynamic thermal effect in the cubic phases: Thermal motion creates redshift in oxide perovskites and blueshift in halide perovskites, whereas the intrinsic local symmetry-breaking usually results in band gap blueshift in oxide and halide perovskites. The observed band gap renormalization in cubic phase at high temperature is then constructive (as in halide perovskite) or destructive (as in oxide perovskites) superposition of the DOWPs effect and thermal motion. Consequently, the real band gap renormalization at finite temperature (T > $T_c$) is underestimated for oxide perovskites and overestimated for halide perovskites with respect to the nominal cubic structure (Pm-3m).

This study clarifies the role of intrinsic symmetry breaking on the properties of cubic perovskites, separating them from the traditional thermal effects. Intrinsic DOWPs rather than thermal motions can break the centrosymmetric symmetry (Pm-3m), making materials acquire unique functionality in cubic phase such as strong Rashba effect, Raman effect, piezoelectric effect or second harmonic generation, etc.

## Acknowledgments

The work on the molecular dynamics and effective band structure at the University of Colorado at Boulder was supported by the DMREF program of the U.S. National Science Foundation through Grant No. DMREF-1921949. The relevant computations were done at the NSF XSEDE supercomputers. Work on the structural optimizations of intrinsic polymorphous networks was supported by the U.S. Department of Energy, Office of Science, Basic Energy Sciences, Materials Sciences and Engineering Division, under Grant No. DE-SC0010467 to the University of Colorado. These ab initio calculations about intrinsic polymorphous network were performed using resources of the National Energy Scientific Computing Center, which is supported by the Office of Science of the U.S. Department of Energy.

We thank Dr. C. M. Culbertson, who kindly shared experimental PDF data of cubic SrTiO$_3$, and thank Dr. M. Zacharias, who kindly shared the trajectory files of MD simulations for cubic SrTiO$_3$.

**References**

[1] M. Yashima, M. Tanaka, J. Appl. Crystallogr. 37 (2004) 786–790.
[2] S. Aoyagi et al., J. Phys. Soc. Jpn. 71 (2002) 1218–1221.
[3] Z. Salman et al., Phys. Rev. Lett. 96 (2006) 147601.
[4] B. Wang, N. Novendra, A. Navrotsky, J. Am. Chem. Soc. 141 (2019) 14501– 14504.
[5] K. Yamada et al., Chem. Lett. 20 (1991) 801–804.
[6] H.M. Rietveld, J. Appl. Crystallogr. 2 (1969) 65–71.
[7] (IUCr) International Tables for Crystallography (accessed Oct 29, 2020).
[8] D. Zagorac et al., J. Appl. Crystallogr. 52 (2019) 918–925.
[9] M.J. Mehl et al., Comput. Mater. Sci. 136 (2017) S1–S828.
[10] A. Jain et al., APL Mater. 1 (2013) 011002.
[11] S. Tariq et al., AIP Adv. 5 (2015) 077111.
[12] S. Saha, T.P. Sinha, A. Mookerjee, Eur. Phys. J. B 18 (2000) 207–214.
[13] S. Saha, T.P. Sinha, A. Mookerjee, Phys. Rev. B 62 (2000) 8828–8834.
[14] R.A. Evarestov, V.P. Smirnov, D.E. Usvyat, Solid State Commun. 127 (2003) 423–426.
[15] A.R. Benrekia et al., Physica B 407 (2012) 2632–2636.
[16] R. Ahuja, O. Eriksson, B. Johansson, J. Appl. Phys. 90 (2001) 1854–1859.
[17] M. Afsari, A. Boochani, M. Hantezadeh, Optik 127 (2016) 11433–11443.
[18] S. McKechnie et al., Phys. Rev. B 98 (2018) 085108.
[19] S. Poncé, M. Schlipf, F. Giustino, ACS Energy Lett. 4 (2019) 456–463.
[20] S. Tinte et al., J. Phys.: Condens. Matter 11 (1999) 9679–9690.
[21] K. Parlinski, Y. Kawazoe, Y. Waseda, J. Chem. Phys. 114 (2001) 2395–2400.
[22] J.-J. Zhou, O. Hellman, M. Bernardi, Phys. Rev. Lett. 121 (2018) 226603.
[23] A. Marronnier, G. Roma, S. Boyer-Richard, L. Pedesseau, J.-M. Jancu, van Bonnassieux, C. Katan, C.C. Stoumpos, M.G. Kanatzidis, J. Even, in: 2018 IEEE 7th World Conference on Photovoltaic Energy Conversion (WCPEC) (A Joint Conference of 45th IEEE PVSC, 28th PVSEC 34th EU PVSEC), 2018, pp. 1715– 1717.
[24] A. Bano et al., AIP Conf. Proc. 1731 (2016) 090001.
[25] R.X. Yang et al., J. Phys. Chem. Lett. 8 (2017) 4720–4726.
[26] R.X. Yang et al., J. Chem. Phys. 152 (2020) 024703.
[27] Z. Wang et al., Phys. Rev. B 102 (2020) 235121.
[28] X.-G. Zhao et al., Phys. Rev. B 101 (2020) 155137.
[29] Z. Wang et al., Physical Review B 103 (2021) 165110.
[30] J. Varignon, M. Bibes, A. Zunger, Physical Review B 100 (2019) 035119.
[31] O.I. Malyi, A. Zunger, Appl. Phys. Rev. 7 (2020) 041310.
[32] J. Varignon, M. Bibes, A. Zunger, Nat. Commun. 10 (2019) 1–11.
[33] Q. Zhang, T. Cagin, W.A. Goddard, PNAS 103 (2006) 14695–14700.
[34] B. Monserrat, D. Vanderbilt, ArXiv:1711.06274 [Cond-Mat] (2017).
[35] W. Prusseit-Elffroth, F. Schwabl, Appl. Phys. A 51 (1990) 361–368.
[36] J.H. Lee, K.M. Rabe, Phys. Rev. B 84 (2011) 104440.
[37] M.-C. Jung, K.-W. Lee, Phys. Rev. B 90 (2014) 045120.
[38] J. Wiktor, U. Rothlisberger, A. Pasquarello, J. Phys. Chem. Lett. 8 (2017) 5507– 5512.
[39] S.R. Phillpot, V. Gopalan, Appl. Phys. Lett. 84 (2004) 1916–1918.

[40] K.A. Müller et al., J. Phyique Lett. 43 (1982) 537–542.
[41] E.A. Stern, Phys. Rev. Lett. 93 (2004) 037601.
[42] W.P. Mason, B.T. Matthias, Phys. Rev. 74 (1948) 1622–1636.
[43] M.E. Lines, A.M. Glass, Principles and Applications of Ferroelectrics and Related Materials, OUP Oxford, 2001.
[44] E.K.H. Salje et al., Phys. Rev. B 87 (2013) 014106.
[45] R. Allmann, R. Hinek, Acta Crystallogr., Sect. A 63 (2007) 412–417.
[46] H.T. Dang, A.J. Millis, C.A. Marianetti, Phys. Rev. B 89 (2014) 161113.
[47] K. Held et al., Int. J. Mod Phys B 15 (2001) 2611–2625.
[48] A. Georges et al., Rev. Mod. Phys. 68 (1996) 13–125.
[49] L. Nordheim, Ann. Phys. 401 (1931) 607–640.
[50] F. Yonezawa, K. Morigaki, Progr. Theor. Phys. Suppl. 53 (1973) 1–76.
[51] A.M. Glazer, Acta Crystallogr., B 28 (1972) 3384–3392.
[52] S. Ishiwata et al., Phys. Rev. B 84 (2011) 054427.
[53] M. Zacharias, M. Scheffler, C. Carbogno, Phys. Rev. B 102 (2020) 045126.
[54] L. Granlund, S.J.L. Billinge, P.M. Duxbury, Acta Crystallogr., Sect. A 71 (2015) 392–409.
[55] K. Sato et al., Phys. Scr. 2005 (2005) 359.
[56] D.C. Arnold et al., Adv. Funct. Mater. 20 (2010) 2116–2123.
[57] J. Sun, A. Ruzsinszky, J.P. Perdew, Phys. Rev. Lett. 115 (2015) 036402.
[58] J.P. Perdew et al., Phys. Rev. Lett. 100 (2008) 136406.
[59] F.W. Lytle, J. Appl. Phys. 35 (1964) 2212–2215.
[60] P.M. Woodward, Acta Crystallogr., B 53 (1997) 32–43.
[61] D.H. Fabini et al., J. Am. Chem. Soc. 138 (2016) (1832) 11820–11821.
[62] R.C. Remsing, M.L. Klein, APL Mater. 8 (2020) 050902.
[63] B.B. Van Aken et al., Nat. Mater. 3 (2004) 164–170.
[64] P.M. Woodward, Acta Crystallogr., Sect. B 53 (1997) 44–66.
[65] J.M. Rondinelli, C.J. Fennie, Adv. Mater. 24 (2012) 1961–1968.
[66] E.H. Smith, N.A. Benedek, C.J. Fennie, Inorg. Chem. 54 (2015) 8536–8543.
[67] P. Garcia-Fernandez et al., J. Phys. Chem. Lett. 1 (2010) 647–651.
[68] A. Mercy et al., Nat. Commun. 8 (2017) 1677.
[69] G. Shirane, H. Danner, R. Pepinsky, Phys. Rev. 105 (1957) 856–860.
[70] M.S. Senn et al., Phys. Rev. Lett. 116 (2016) 207602.
[71] C.M. Culbertson et al., Sci. Rep. 10 (2020) 1–10.
[72] A. Belsky et al., Acta Crystallogr., B 58 (2002) 364–369.
[73] A.N. Beecher et al., ACS Energy Lett. 1 (2016) 880–887.
[74] X. Wei, G. Xu, Z. Ren, C. Xu, W. Weng, G. Shen, G. Han, J. Am. Ceram. Soc. 93 (2010) 1297–1305.
[75] V. Popescu, A. Zunger, Phys. Rev. Lett. 104 (2010) 236403.
[76] G.M. Dalpian et al., Chem. Mater. 31 (2019) 2497–2506.
[77] Z. Xiao, W. Meng, J. Wang, D. B. Mitzi, and Y. Yan. Materials Horizons 2 (2017) 206-216.
[78] A. Tkatchenko, M. Scheffler, Phys. Rev. Lett. 102 (2009) 073005.
[79] M. De Raychaudhury, E. Pavarini, O.K. Andersen, Phys. Rev. Lett. 99 (2007) 126402.